\documentclass[
reprint,
superscriptaddress,
amsmath,amssymb,
prl,
longbibliography
]{revtex4-2}

\usepackage{graphicx}
\usepackage{dcolumn}
\usepackage{bm}
\usepackage[mathlines]{lineno}
\usepackage{mathtools}
\usepackage{amsthm}
\usepackage{multirow}
\usepackage{array}
\usepackage{makecell}
\usepackage{physics}

\usepackage{hyperref}
\hypersetup{colorlinks=true, linkcolor=blue, urlcolor=blue, citecolor=blue, pdfstartview={FitH}, hyperfootnotes=false, unicode=true}

\newtheorem{theorem}{Theorem}
\newtheorem{corollary}{Corollary}

\def\be{\begin{equation}}
\def\ee{\end{equation}}

\newcommand{\Ncal}{\mathcal{N}}
\newcommand{\Ccal}{\mathcal{C}}

\begin{document}

\title{Uncertainty Principle from Operator Asymmetry}

\author{Xingze Qiu}
\email{xingze@tongji.edu.cn}
\affiliation{School of Physics Science and Engineering, Tongji University, Shanghai 200092, China}

\date{\today}

\begin{abstract}

The uncertainty principle is fundamentally rooted in the algebraic asymmetry between observables. We introduce a new class of uncertainty relations grounded in the resource theory of asymmetry, where incompatibility is quantified by an observable's intrinsic, state-independent capacity to break the symmetry associated with another. This ``operator asymmetry,'' formalized as the incompatibility norm, leads to a variance-based uncertainty relation for pure states that can be tighter than the standard Robertson bound. Most significantly, this framework resolves a long-standing open problem in quantum information theory: the formulation of a universally valid, product-form uncertainty relation for the Wigner-Yanase skew information. We demonstrate the practical power of our framework by deriving tighter quantum speed limits for the dynamics of nearly conserved quantities, which are crucial for understanding non-equilibrium phenomena such as prethermalization and many-body localization. This work provides both a new conceptual lens for understanding quantum uncertainty and a powerful, versatile toolkit for its application.

\end{abstract}

\maketitle

%%%%%%%%%%%%%%%%%%%%%%%%%%%%%%%%

{\it Introduction.---}
%%%%%%%%%%%%%%%%%%%%%%%%%%%%%%%%
The uncertainty principle, first articulated by Heisenberg \cite{Heisenberg_1927} and formalized by Kennard \cite{Kennard_1927} and Robertson \cite{Robertson_1929_PhysRev}, represents a fundamental departure from classical physics by asserting that the statistical distributions of measurement outcomes for certain pairs of observables cannot be simultaneously sharp in a given quantum state. In its modern form, the Robertson uncertainty relation for two non-commuting observables $A$ and $B$ is given by \cite{Sakurai_Napolitano_2020} 
\be
\Delta A\cdot \Delta B \geq \frac{1}{2} \abs{\expval{[A,B]}}\, .
\label{eq:Robertson_intro}
\ee
Here, $\Delta A = \sqrt{\langle A^2\rangle - \langle A\rangle^2}$ is the standard deviation of $A$ in a given quantum state $\rho$, and $\expval{\cdot} = \Tr{\rho \cdot}$ denotes the expectation value. 
Despite its foundational importance, this relation suffers from a critical flaw: the state-dependent lower bound can vanish for certain states even if the observables do not commute, rendering it trivial in many physical scenarios. This limitation has spurred decades of research into more robust formulations, leading to significant advances such as entropic uncertainty relations, which provide state-independent bounds \cite{Deutsch_1983, Maassen_1988, Berta_2010, Coles_2017_RMP}, and majorization-based relations that offer a more fine-grained description of uncertainty \cite{Portesi_2006, Puchala_2013, Partovi_2011, Friedland_2013, Portesi_2016}. 
Another important direction has focused on information-theoretic measures of uncertainty for mixed states, most notably the Wigner-Yanase skew information (WYSI) \cite{Luo_2003_PRL, Luo_2004_skew, Luo_2005_PRA, Kosaki_2005, Rivas_2008_PRA, Li_2009_PRA, Chen_2016_skew, Huang_2020_skew, Cai_2021_skew, Kenjiro_2010_gWYD, Fan_2018_gWYD, Wu_2020_gWYD, Yang_2022_gWYSI, Frank_2008_PNAS, Ren_2021_PRA, Hu_2023, Laurent_2024}, which quantifies the uniquely quantum component of uncertainty. However, the quest for a general, product-form uncertainty relation for WYSI has been a long and challenging one \cite{Laurent_2024}.

\begin{figure}[tp!]
\centering
\includegraphics[width=0.48\textwidth]{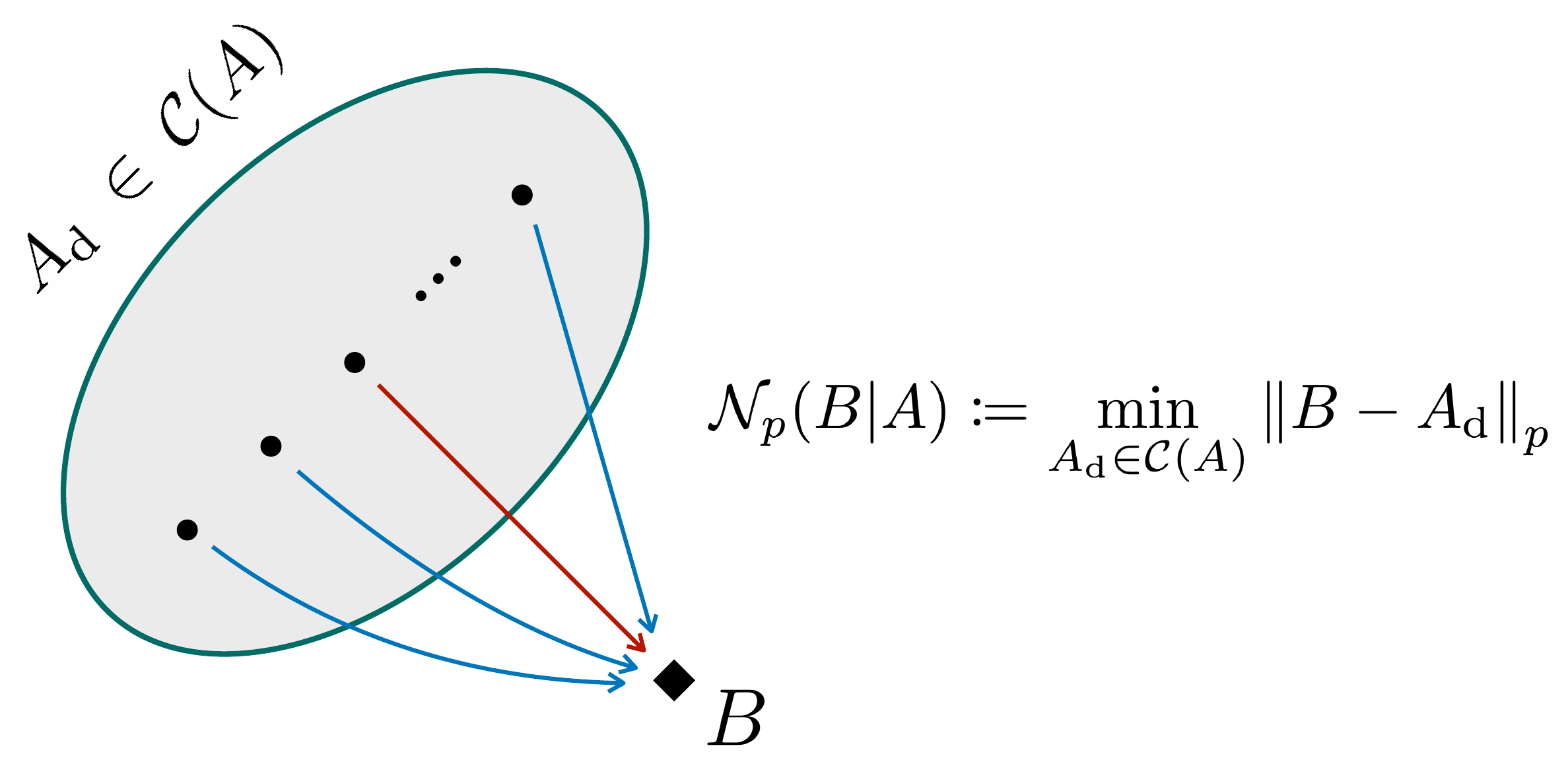}
\caption{Schematic interpretation of the incompatibility norm. The set of operators that commute with observable $A$ forms a commutant algebra, $\Ccal(A)$ (oval), representing all operators (black dots) compatible with $A$. The incompatibility of another observable $B$ (black diamond) is quantified by its minimal distance (red arrow) to this algebra. This distance, the incompatibility norm $\Ncal_p(B|A)$ [Eq.~\eqref{eq:IncompatibilityNorm_definition}], measures the magnitude of the essential symmetry-breaking component of $B$ relative to $A$. This state-independent quantity forms the foundation of our asymmetry-bounded uncertainty relations. 
}
\label{fig:Fig_1}
\end{figure}

In this work, we introduce a new paradigm that reframes quantum incompatibility itself. Instead of focusing on the state-dependent expectation of the commutator, we quantify the intrinsic, state-independent structural incompatibility between observables. This approach is deeply rooted in the operational framework of quantum resource theories \cite{Chitambar_2019_RMP, Gour_2025}, specifically the resource theory of asymmetry (RTA). In RTA, asymmetry is a resource that enables the breaking of a physical symmetry \cite{Marvian_2014_NC, Marvian_2016_PRA, Takagi_2022_PRL, Yamaguchi_2023_PRL}. We extend this logic from states to operators: the incompatibility of an observable $B$ with respect to $A$ is precisely its capacity to act as a resource for breaking the symmetry associated with $A$. As depicted in Fig.~\ref{fig:Fig_1}, we formalize this concept as the incompatibility norm, a state-independent measure of incompatibility.

\begin{table*}
    \centering
    \caption{
    A comparison between our asymmetry-bounded uncertainty relations and established uncertainty relations for variance and Wigner-Yanase Skew Information (WYSI), respectively. Here, $\mathcal{R}$ (Cor.~\ref{Cor_I}) and $\widetilde{\mathcal{R}}$ (Cor.~\ref{Cor_II}) are the factors defined in the respective corollaries, $C = -i[A,B]$, $\Delta O$ is the standard deviation of observable $O$ in a pure state $\ket{\psi}$, $\expval{C} = \expval{\psi|C|\psi}$, and $I(\rho,O)$ is the WYSI of a state $\rho$ with respect to $O$ [Eq.~\eqref{eq:wysi_def}].
    }
    \vspace{5pt}
    \renewcommand\arraystretch{2.5}
    \resizebox{1\linewidth}{!}{
        \begin{tabular}{  l  | c | c }
        \hline
        \hline
        
        \makecell[c] \; Uncertainty Measure \;  & Asymmetry-Bounded Uncertainty Relations (This Work) &  Established Uncertainty Relations  \\
        \hline
        
        \makecell[c]{ Variance (Pure State)}
       
        & {\makecell[c]{ 
         {$\displaystyle \Delta A\cdot \Delta B \geq \frac{1}{2} \abs{\expval{C}}\cdot \mathcal{R} $}\quad [Eq.~\eqref{eq:pure_state_AUR_main}] \\
         }}
         
        & {$\displaystyle \Delta A\cdot \Delta B \geq \frac{1}{2} \abs{\expval{C}}  $} \quad [Eq.~\eqref{eq:Robertson_intro}] \\[8pt] 

        \hline
        \makecell[c]{ WYSI (Mixed State)}
        
        & {\makecell[c]{ 
         {$\displaystyle \sqrt{I(\rho,A)} \cdot \sqrt{I(\rho,B)} \geq \frac{1}{2}\abs{\Tr{\sqrt{\rho} C}}\cdot \widetilde{\mathcal{R}} $} \quad [Eq.~\eqref{eq:wysi_AUR_main}] \\ 
         }}
         
        & {$\displaystyle \sqrt{I(\rho,A)} \cdot \sqrt{I(\rho,B)} \geq \frac{1}{2}\abs{\Tr{\rho C}} $} \quad (See \cite{Luo_2003_PRL}) \\[8pt] 
          
        \hline 
        \hline
        \end{tabular}
        }
    \label{tab:compare_AUR}
\end{table*}

This framework yields a new class of asymmetry-bounded uncertainty relations (AURs). Here, ``bounded'' emphasizes that the uncertainty product is constrained from below by a quantity determined by intrinsic quantum incompatibility. Our main contributions are threefold. First, we derive two families of general AURs (Thm.~\ref{Thm:AUR_I} and Thm.~\ref{Thm:AUR_II}). Second, as direct consequences, we obtain a Robertson-type variance-based uncertainty relation for pure states that can be significantly tighter than the standard bound (Cor.~\ref{Cor_I}), and a universally valid product-form uncertainty relation for the WYSI, which resolves a major open problem in quantum information theory (Cor.~\ref{Cor_II}). These results are summarized in Tab.~\ref{tab:compare_AUR}. Third, we demonstrate the practical utility of our framework by deriving tighter quantum speed limits (QSLs) [Eq.~\eqref{eq:QSLs}] for the dynamics of nearly conserved quantities, a crucial element in the study of non-equilibrium many-body systems. 
We also expand the horizon of this framework, detailing its applications in entanglement detection, quantum metrology, and quantum thermodynamics. 
Our work thus unifies concepts from quantum foundations, information theory, and many-body physics, providing a new perspective through which to view and quantify one of the most fundamental principles of quantum theory.

%%%%%%%%%%%%%%%%%%%%%%%%%%%%%%%%

{\it Quantifying incompatibility via operator asymmetry.---}
%%%%%%%%%%%%%%%%%%%%%%%%%%%%%%%%
Our framework is built upon RTA. For an observable $A$, its commutant algebra, $\Ccal(A)$, is the set of all operators that commute with $A$. In the language of resource theories \cite{Chitambar_2019_RMP}, these operators can be considered as the ``free'' ones with respect to the eigenbasis of $A$. The degree to which another observable, $B$, is incompatible with $A$ can be quantified by its minimal distance to this set of free operators. This motivates our central definition for quantifying incompatibility. Inspired by the resource-theoretic quantification of asymmetry \cite{Chitambar_2019_RMP}, we define the incompatibility norm of $B$ with respect to $A$ as
\be
    \Ncal_p(B|A) \coloneqq \min_{A_{\rm d} \in \Ccal(A)} \norm{B - A_{\rm d}}_p\, .
    \label{eq:IncompatibilityNorm_definition}
\ee
Here, $\norm{P}_p \coloneqq (\Tr{\abs{P}^p})^{1/p}$ is the Schatten $p$-norm with $\abs{P}=\sqrt{P^\dag P}$ \cite{Bhatia_2013_matrix}. 
As shown in Fig.~\ref{fig:Fig_1}, this quantity represents the minimal distance from $B$ to the algebra of operators compatible with $A$. The minimization procedure effectively decomposes $B$ into a ``free'' part $A_{\rm d} \in \Ccal(A)$ and a ``resourceful'' or symmetry-breaking part $B - A_{\rm d}$. The incompatibility norm $\Ncal_p(B|A)$ quantifies the minimal ``size'' of this resourceful component, which is responsible for generating asymmetry relative to $A$. This quantity is state-independent and provides a fundamental measure of the incompatibility between $A$ and $B$. 
The term ``operator asymmetry'' is used to provide physical intuition for this concept, representing the intrinsic power of operator $B$ to induce asymmetry relative to $A$. Our terminology ``incompatibility norm'' is justified by its structural analogy to asymmetry measures for quantum states, which quantify a state's distance to the set of symmetric (free) states. Here, we extend this logic to quantify the asymmetry of operators.

It is crucial to distinguish the scope of our framework, which addresses preparation uncertainty (intrinsic statistical constraints) \cite{Ballentine_1970_RMP}, from measurement uncertainty (device error and disturbance) \cite{Busch_2014_RMP, Mao_2023_PRL}. 
While both arise from non-commutativity, they represent distinct physical manifestations of the uncertainty principle.
Unlike measurement uncertainty approaches that typically minimize distances between probability distributions to quantify device performance of a measurement apparatus, our incompatibility norm operates directly on the algebraic structure of observables (Hermitian operators). This structural distinction highlights our framework as a unique paradigm grounded in the RTA, providing a state-independent measure of the intrinsic algebraic incompatibility underlying preparation uncertainty.

%%%%%%%%%%%%%%%%%%%%%%%%%%%%%%%%

{\it Asymmetry-bounded uncertainty relations.---}
%%%%%%%%%%%%%%%%%%%%%%%%%%%%%%%%
The derivation of our AURs begins with the standard identity underlying the Robertson relation. For the commutator $[A,B]=iC$, we have $i\Tr{\rho C} = \Tr{\rho[A,B]}$. 
The crucial step is to recognize that for any $A_{\rm d} \in \Ccal(A)$, the commutator remains unchanged: $[A,B] = [A, B-A_{\rm d}]$ since $[A, A_{\rm d}]=0$. This allows us to write $\Tr{\rho C} = -i\Tr{\rho[A, B-A_{\rm d}]} = -i\Tr{[\rho,A](B-A_{\rm d})}$, where we have used the cyclic property of trace. 
Applying H\"older's inequality for conjugate exponents $p, \, q \in [1, \infty]$ satisfying $1/p + 1/q = 1$ \cite{Bhatia_2013_matrix}, we obtain: 
\be
    \abs{\Tr{\rho C}} \leq \norm{B-A_{\rm d}}_p \cdot  \norm{[\rho, A]}_q \, .
    \label{eq:holder_step}
\ee
Since this holds for any $A_{\rm d} \in \Ccal(A)$, we can select the one minimizing the right-hand side, yielding the tightest possible bound. 
This by definition introduces the incompatibility norm: $\abs{\Tr{\rho C}} \leq \Ncal_p(B|A) \cdot \norm{[\rho,A]}_q$. 
By symmetry, we also have $\abs{\Tr{\rho C}} \leq \Ncal_r(A|B) \cdot \norm{[\rho,B]}_s$ for conjugate exponents $r, \, s \in [1, \infty]$ with $1/r + 1/s = 1$, where $\Ncal_r(A|B) \coloneqq \min_{B_{\rm d} \in \Ccal(B)} \norm{A - B_{\rm d}}_r$. Multiplying these two inequalities leads to our first main result.

\begin{theorem}[General AURs: I]\label{Thm:AUR_I}
For any observables $A, B$, quantum state $\rho$, and pairs of conjugate exponents $(p,\, q)$ and $(r,\, s)$, the following uncertainty relation holds:
\begin{equation}
    \norm{[\rho,A]}_q \cdot \norm{[\rho,B]}_s \geq \frac{\abs{\Tr{\rho C}}^2}{\Ncal_p(B|A) \cdot \Ncal_r(A|B)}\, ,
    \label{eq:AUR_theorem_main}
\end{equation}
where $C = -i[A,B]$, $\Ncal_p(B|A)$ and $\Ncal_r(A|B)$ are the corresponding incompatibility norms. 
\end{theorem}

These relations recast the uncertainty bound not in terms of the commutator's expectation value alone, but as a ratio of this value to the algebraic incompatibility of the observables.

The above general AURs can be specialized to a more familiar form involving variances for pure states. 
For $\rho = \ket{\psi}\!\bra{\psi}$, the singular values of $[\rho, A]$ have exactly two non-zero values, both equal to $\Delta A$. Therefore, the Schatten $q$-norm of the commutator with the state relates directly to the variance: $\norm{[\rho,A]}_q = 2^{1/q} \Delta A$. 
Substituting this identity for both $A$ and $B$ into Eq.~\eqref{eq:AUR_theorem_main}, we arrive at a corollary for pure states, which provides a distinct and robust bound, particularly in the regime where the Robertson relation becomes trivial. 

\begin{corollary}[Pure State AURs for Variances]\label{Cor_I}
For a pure state $\ket{\psi}$ and any pairs of conjugate exponents $(p,\,q)$ and $(r,\, s)$,
\begin{equation}
    \Delta A \cdot \Delta B \geq \frac{1}{2} \abs{\expval{C}} \cdot \mathcal{R}\, ,
    \label{eq:pure_state_AUR_main}
\end{equation}
where the factor is given by
\begin{equation}
    \mathcal{R} = 2^{1-(1/q+1/s)} \frac{\abs{\expval{C}}}{\Ncal_p(B|A) \cdot \Ncal_r(A|B)}\, .
    \label{eq:R_factor_AUR}
\end{equation}
\end{corollary}

It is important to emphasize that while Eq.~\eqref{eq:pure_state_AUR_main} can be cast as a refinement of the Robertson relation, it is fundamentally a distinct constraint. 
Specifically, in the near-compatible limit where the commutator vanishes ($|\langle C \rangle| \to 0$), the factor $\mathcal{R}$ diverges, precisely compensating for the vanishing Robertson bound. This yields a finite, non-trivial constraint determined by the intrinsic algebraic incompatibility, ensuring the bound remains robust even when the standard relation fails. 
The AUR thus provides a robust constraint that survives in regimes where the Robertson bound becomes trivial.

We now provide a fully analytical example demonstrating the AUR's superiority over the Robertson bound. 
Consider a qubit with observables: $A = \sigma_z$ and $B(\theta) = \cos(\theta)\sigma_z + \sin(\theta)\sigma_x$ for $\theta \in [0, \pi/2]$, giving the commutator $C(\theta) = -i[A,B(\theta)] = 2\sin(\theta)\sigma_y$. Here, $\sigma_{x,y,z}$ are the usual Pauli operators. 
The Robertson bound is $\Delta A \cdot \Delta B \geq  \abs{\expval{\psi|\sigma_y|\psi}} \sin(\theta)$, which vanishes as the observables become compatible ($\theta \to 0$). 
To analytically calculate the AUR bound, we first find the incompatibility norms. 
For $\Ncal_p(B(\theta)|A)$, the commutant algebra $\Ccal(A)$ consists of operators that are diagonal in the $z$-basis. The optimal choice for $A_{\rm d} \in \Ccal(A)$ is the projection of $B(\theta)$ onto $\Ccal(A)$, which is $A_{\rm d} = \cos(\theta)\sigma_z$. Thus, $\Ncal_p(B(\theta)|A) = \norm{\sigma_{x}}_{p} \sin(\theta) = 2^{1/p} \sin(\theta) $. 
By symmetry, a straightforward calculation in the eigenbasis of $B(\theta)$ yields $\mathcal{N}_{r}(A|B(\theta)) = 2^{1/r} \sin(\theta)$. 
Substituting these exact results into Eq.~\eqref{eq:R_factor_AUR} gives the factor $\mathcal{R} = \abs{\expval{\psi|\sigma_y|\psi}}/\sin(\theta)$. 
From Eq.~\eqref{eq:pure_state_AUR_main}, we obtain the AUR bound $\Delta A \cdot \Delta B \geq \abs{\expval{\psi|\sigma_y|\psi}}^2$. 
This result is striking: as $\theta \to 0$, the Robertson bound vanishes linearly, whereas the AUR bound remains constant and strong, demonstrating its power in the near-compatible regime where the standard relation is most vulnerable.

%%%%%%%%%%%%%%%%%%%%%%%%%%%%%%%%

{\it Information-theoretic AURs: a solution to the WYSI problem.---}
%%%%%%%%%%%%%%%%%%%%%%%%%%%%%%%%
The true power of our framework is most profoundly demonstrated by its ability to solve a major open problem in quantum information theory: the formulation of a general, product-form uncertainty relation for the WYSI \cite{Wigner_Yanase_1963}: 
\begin{equation}
    I(\rho,A) = -\frac{1}{2}\Tr{[\sqrt{\rho}, A]^2} = \frac{1}{2} \norm{[\sqrt{\rho}, A]}_2^2\, .
    \label{eq:wysi_def}
\end{equation}
It represents the amount of information the state $\rho$ contains about observables not commuting with (being skew to) $A$ \cite{Luo_2003_PRL}. 
For pure states, WYSI reduces to the variance: $I(\ket{\psi}\!\bra{\psi}, A) = \Delta A^2$. 
For general mixed states, WYSI is of fundamental importance as it quantifies the uniquely quantum component of an observable's uncertainty, excluding contributions from classical mixing \cite{Luo_2006_PRA_uncertainty}. 
WYSI is also convex with respect to the state \cite{Lieb_1973_PRL, Lieb_1973_convexity, Wehrl_1978_RMP} and is additive for local observables \cite{Luo_2003_PRL}, which are crucial for any measure of quantum information. 
In addition, its deep connection to quantum Fisher information (QFI) \cite{Luo_2003_PRL, Yang_2022_gWYSI, Pires_2016_PRX} makes it central to quantum metrology and measurement theory \cite{Helstrom_1969, Braunstein_1994_PRL, Toth_2012_PRA}.

WYSI, as a more refined measure of quantum uncertainty, provides a powerful foundation for developing new and often stronger uncertainty relations. 
An early, influential product-form relation was proposed by Luo \cite{Luo_2003_PRL}: $\sqrt{I(\rho,A)} \cdot \sqrt{I(\rho,B)} \geq \abs{\Tr{\rho C}}/2$, which promised a powerful new tool. 
However, this relation was later shown to be not generally valid through explicit counterexamples \cite{Kosaki_2005, Rivas_2008_PRA}, revealing unexpected subtleties in its derivation \cite{Li_2009_PRA}. This discovery largely shifted the field's focus towards sum-form uncertainty relations, which bound quantities like $I(\rho,A) + I(\rho,B)$ \cite{Chen_2016_skew, Huang_2020_skew, Cai_2021_skew}. 
Other studies have focused on generalizations such as the Wigner-Yanase-Dyson skew information \cite{Kenjiro_2010_gWYD, Fan_2018_gWYD, Wu_2020_gWYD, Yang_2022_gWYSI} and the more general metric-adjusted skew information \cite{Frank_2008_PNAS, Ren_2021_PRA, Hu_2023}. 
While valuable, these sum-form relations do not capture the reciprocal trade-off inherent in the Heisenberg principle, nor do they serve as a direct product-form analogue to the Robertson relation for WYSI. 
Crucially, the product form is uniquely necessary to quantify how high precision (information) in one observable inevitably constrains the precision of its non-commuting partner. 
The subtleties uncovered in the study of Luo's pioneering relation have revealed that the geometric structure of quantum uncertainty for mixed states is far more intricate than for pure states. 
The recent breakthrough for the canonical position-momentum pair, achieved through the framework of quantum Sobolev inequalities, represents the current frontier \cite{Laurent_2024}. 
However, the formulation of a robust, general product-form uncertainty relation for the original WYSI that applies to any pair of non-commuting observables has remained an open challenge.

Our asymmetry framework provides a natural and direct solution. 
The derivation mirrors that of Thm.~\ref{Thm:AUR_I}, with the state $\rho$ replaced by its square root, $\sqrt{\rho}$. Starting from $\Tr{\sqrt{\rho}C} = -i \Tr{[\sqrt{\rho},A](B-A_{\rm d})}$ and applying H\"older's inequality leads to our second main result.

\begin{theorem}[General AURs: II]\label{Thm:AUR_II}
For any observables $A, B$, state $\rho$, and conjugate exponents $(p,q)$ and $(s,t)$, the following uncertainty relation holds:
\be
    \norm{[\sqrt{\rho},A]}_q \cdot \norm{[\sqrt{\rho},B]}_s \geq \frac{\abs{\Tr{\sqrt{\rho} C}}^2}{\Ncal_p(B|A) \cdot \Ncal_r(A|B)}\, .
\ee
\end{theorem}

For pure states, this theorem coincides with Thm.~\ref{Thm:AUR_I}. 
For general mixed states, this theorem elegantly resolves the long-standing problem. 
By setting $p=r=2$ (so $q=s=2$) and using the definition in Eq.~\eqref{eq:wysi_def}, we obtain a direct product-form uncertainty relation for the WYSI.

\begin{corollary}[Product-Form WYSI-AUR]\label{Cor_II}
\begin{equation}
    \sqrt{I(\rho,A)} \cdot \sqrt{I(\rho,B)} \geq \frac{1}{2}\abs{\Tr{\sqrt{\rho} C}}\cdot \widetilde{\mathcal{R}}\, ,
    \label{eq:wysi_AUR_main}
\end{equation}
where $\widetilde{\mathcal{R}} =  \abs{\Tr{\sqrt{\rho} C}}\cdot[\Ncal_2(B|A) \cdot \Ncal_2(A|B)]^{-1}$. 
\end{corollary}

This relation is universally valid. 
To demonstrate its validity where previous attempts failed, we consider the well-known counterexample discussed in the literature: $A=\sigma_x$, $B=\sigma_y$, and $\rho = {\rm diag}\{0.75,0.25\}$. Here, $C=2\sigma_z$. A direct calculation yields $I(\rho,A)=I(\rho,B)=[(\sqrt{3}-1)/2]^2\approx 0.134$ and $\abs{\Tr{\rho C}}=1$. The inequality $\sqrt{I(\rho,A)} \cdot \sqrt{I(\rho,B)}\approx 0.134\geq 0.5$, suggested by Luo's relation, does not hold here. 
For our relation, we calculate $\Ncal_2(B|A)=\Ncal_2(A|B)=\sqrt{2}$ and $\abs{\Tr{\sqrt{\rho} C}}=\sqrt{3}-1$. Therefore, the inequality of our relation Eq.~\eqref{eq:wysi_AUR_main} is satisfied precisely as an equality, demonstrating that our relation is not only valid but also tight in this case. 
The algebraic simplicity and generality of this solution complement the sophisticated analytical results recently obtained for specific systems \cite{Laurent_2024}, suggesting that our framework offers a natural perspective on the essential structure of quantum incompatibility.

%%%%%%%%%%%%%%%%%%%%%%%%%%%%%%%%

{\it Tighter quantum speed limits.---}
%%%%%%%%%%%%%%%%%%%%%%%%%%%%%%%%
Our framework has the ability to provide tight, non-trivial bounds and thus is central to many cutting-edge applications in quantum physics. We here present a concrete example demonstrating the AUR's practical advantages to provide tighter QSLs for Observables.

QSLs are direct consequences of time-energy uncertainty relations, setting fundamental bounds on the rate of any physical process \cite{Mandelstam_1945, Deffner_2013_PRL, Deffner_2017_review}. While most often discussed for state evolution, a more practical QSL, first considered by Mandelstam and Tamm \cite{Mandelstam_1945}, bounds the rate of change of an observable's expectation value \cite{Gorshkov_2022_PRX, Hamazaki_2022_PRXQuantum, Mohan_2022_PRA}. For a system evolving under a time-independent Hamiltonian $H$, the velocity $v_A \coloneqq \abs{d\expval{A}/dt}$ of an observable $A$ is given by the Heisenberg equation of motion, $v_A = \abs{\expval{[H,A]}}$. Here, $\expval{\cdot}=\Tr{\rho_t \cdot}$, $\rho_t=U_t\rho_0 U^\dag_t$, $U_t=\exp(-iHt)$, and $\rho_0$ is the initial state. 
Applying the Robertson uncertainty relation to the commutator yields the standard Mandelstam-Tamm QSL for an observable: $v_A \leq 2 \Delta A \cdot \Delta H$. 
This bound, however, becomes trivially loose for nearly conserved quantities, which are central to understanding transport in many-body systems \cite{Hamazaki_2024_PRR}. 
For such an observable, $[H, A]$ is a ``small'' operator (its norm is small), and thus its expectation value, i.e., the velocity $v_A$, is also small: $v_A = \abs{\expval{[H,A]}}\ll 1$. 
Yet, the variances $\Delta A$ and $\Delta H$ can still be very large, making the standard bound uninformative, thereby failing to capture the true physical constraint on the evolution timescale.

Our framework rectifies this failure. Instead of applying the Robertson relation, we utilize the same logic as in Thm.~\ref{Thm:AUR_I}, which gives a AUR-like bound: $v^2_A \leq \norm{[\rho, H]}_q\cdot \norm{[\rho, A]}_s \cdot \Ncal_p(A|H) \cdot \Ncal_r(H|A)$ with conjugate exponents $(p,q)$ and $(s,t)$. 
Specializing to a pure initial state $\rho_0 = \ket{\psi_0}\!\bra{\psi_0}$ and the most common case of the Schatten 2-norm ($p=q=2$, $r=s=2$), we have $\norm{[\rho_t, H]}_2 = \sqrt{2} \Delta H$ and $\norm{[\rho_t, A]}_2 = \sqrt{2} \Delta A$. Substituting these into the inequality immediately leads to our new, AUR-based QSL: 
 \be
v^2_A \leq  2\Delta A\cdot \Delta H \cdot \Ncal_2(A|H) \cdot \Ncal_2(H|A)\, .
\label{eq:QSLs}
\ee
This bound tightens the standard QSL by a factor equal to the product of the incompatibility norms. 
A quantity $A$ is ``nearly conserved'' if it ``almost commutes'' with $H$. Our incompatibility norm $\Ncal_2(A|H)$ provides a rigorous, quantitative measure of this approximate commutation. 
For a nearly-conserved quantity, $A$ is almost diagonal in the energy eigenbasis, meaning $\Ncal_2(A|H)$ is small by definition [by symmetry, $\Ncal_2(H|A)$ is also small]. This factor thus provides the necessary tightening that the standard bound lacks, forging a direct link between the algebraic structure of an approximate conservation law and the physical timescale of its dynamics \cite{Chen_2023, Jonah_2022_Scrambling}. This thus consequently provides a rigorous tool for analyzing the slow dynamics characteristic of phenomena like prethermalization \cite{Mori_2018, Mallayya_2019_PRX, Ueda_2020_NRP} and many-body localization \cite{Nandkishore_2015_Review, Abanin_2019_RMP, Long_2023_PRL}, where evolution is governed by such approximate symmetries.

%%%%%%%%%%%%%%%%%%%%%%%%%%%%%%%%

{\it Conclusion and outlook.---}
%%%%%%%%%%%%%%%%%%%%%%%%%%%%%%%%
By quantifying the incompatibility between two observables via the incompatibility norm---a state-independent measure of an observable's intrinsic power to break a symmetry---we have moved beyond the limitations of the traditional Robertson relation. Our key contributions are the derivation of general AURs and, most significantly, the resolution of a long-standing challenge in quantum information: the formulation of a robust, universally valid product-form uncertainty relation for the WYSI. We have also established tighter and more physically meaningful QSLs for many-body systems with nearly conserved quantities.

The conceptual approach presented here, which connects quantum foundations, information theory, and many-body dynamics, opens several promising avenues for future research: 
(1) {\it Entanglement Detection.} 
For separable states, the sum of local uncertainties is bounded below; a violation of such bounds witnesses entanglement  \cite{Guhne_2004_PRL, Chen_2005_PRA, Guhne_2009_PhysRep}.
Standard variance-based criteria often struggle with mixed states due to classical noise. By utilizing WYSI, which filters out this classical component, our relations offer inherently tighter bounds. 
Crucially, our contribution of a product-form relation for the WYSI allows for the construction of new, more robust witness, which is superior in detecting asymmetric correlations, such as those in squeezed states \cite{Guhne_2009_PhysRep}. This allows our framework to detect entanglement in noisy regimes (e.g., Werner states with low visibility) where variance-based witnesses ofen fail. 
(2) {\it Quantum Metrology.}
Given the intrinsic link between WYSI and the QFI, our universally valid WYSI relation implies fundamental limits for multi-parameter estimation \cite{Giovannetti_2006_PRL, Lu_2021_PRL, Hou_2021_SciAdv}. 
This establishes that precise simultaneous estimation of incompatible parameters is limited not just by the commutator, but by the algebraic asymmetry of the generators, potentially refining the quantum Cram\'er-Rao bound in multi-parameter scenarios. 
(3) {\it Quantum Thermodynamics.}
Asymmetry is a key resource in quantum thermodynamics \cite{Marvian_2022_PRL, Van_2023_PRX, Kondra_2024_PRL, Brandner_2025_PRL}. 
Our formalism directly connects operator asymmetry to the generation of quantum coherence and the thermodynamic costs of precise timekeeping and work extraction. 
Consequently, our framework could be adapted to derive novel thermodynamic uncertainty relations that explicitly link dissipation and fluctuations to the degree of symmetry breaking, bridging the gap between abstract resource theories and practical thermodynamic bounds. 
Ultimately, by grounding uncertainty in the operational language of asymmetry, this work not only resolves a foundational problem but also provides a versatile and rigorous toolkit for exploring the quantum world.

\begin{acknowledgments}
%%%%%%%%%%%%%%%%%%%%%%%%%%%%%%%%

{\it Acknowledgments.---}
%%%%%%%%%%%%%%%%%%%%%%%%%%%%%%%%
We would like to thank Wei Yi and Yunqian Xia for helpful discussions. 
This work is supported by the Fundamental Research Funds for the Central Universities (22120240278) and Shanghai Science and Technology project (24LZ1401600).

\end{acknowledgments}

\bibliography{References}

\end{document}